\documentclass[prd,aps,superscriptaddress,nofootinbib,showpacs,a4paper,notitlepage, 11pt]{revtex4-1} 
\usepackage{times}
\usepackage[utf8]{inputenc}
\usepackage[T1]{fontenc}
\usepackage[UKenglish]{babel}
\usepackage{array}
\usepackage{amsmath}
\usepackage{graphicx}
\usepackage[colorlinks]{hyperref}
\usepackage{latexsym}
\usepackage{amsfonts}
\usepackage{pifont}
\usepackage{stmaryrd}
\usepackage{amssymb}
\usepackage{epsfig}
\usepackage[labelformat=default, labelsep=default, justification=centerlast,  font=small]{caption}
\usepackage{verbatim}
\usepackage{multirow}
\usepackage{rotating}
\usepackage{epstopdf}
\usepackage{geometry}

\newcommand{\be}{\begin{equation}}
\newcommand{\ee}[1]{\label{#1} \end{equation}}
\newcommand{\bbe}{\begin{equation*}}
\newcommand{\eee}{\end{equation*}}
\renewcommand{\arraystretch}{1.3}

\def\f12{\frac{1}{2}}
\newcommand{\bra}{\langle}
\newcommand{\ket}{\rangle}

\begin{document}

\title{Quantum formulation of the Einstein Equivalence Principle}
\author{Magdalena Zych}
\author{\v{C}aslav Brukner}
\affiliation{Faculty of Physics, University of Vienna, Boltzmanngasse 5, A-1090 Vienna, Austria}
\affiliation{Institute for Quantum Optics and Quantum Information, Austrian Academy of Sciences, Boltzmanngasse 3, A-1090 Vienna, Austria.}


\begin{abstract}
Validity of just a few physical conditions comprising the Einstein Equivalence Principle (EEP)  suffices to ensure that gravity can be understood as space-time geometry. EEP is therefore subject to an ongoing experimental verification, with present day tests reaching the regime where quantum mechanics becomes relevant. Here we show that the classical formulation of the EEP does not apply in such a regime. The EEP requires equivalence between the total rest mass-energy of a system, the mass-energy that constitutes its inertia, and the mass-energy that constitutes its weight. In quantum mechanics internal energy is given by a Hamiltonian operator describing dynamics of internal degrees of freedom. We therefore introduce a quantum formulation of the EEP --  equivalence between the rest, inertial and gravitational internal energy \textit{operators}.  We show that the validity of the classical EEP does not imply the validity of its quantum formulation, which thus requires an independent experimental verification. We reanalyse some already completed experiments with respect to the quantum EEP and discuss to which extent they allow testing its various aspects.  
\end{abstract}
\maketitle

\section{Introduction}
\label{sec:intro} 
General relativity describes a  very particular field among other fundamental fields of nature: On one hand,  its dynamics  depends on the mass-energy of matter, on the other,  it also universally governs the dynamics of matter. Whereas the former aspect renders general relativity a theory of gravity, the universality of the field's influence on matter  allows identifying it with the space-time itself, more precisely, with the space-time metric. The importance of the equivalence principle is that it provides conditions, independent of the mathematical framework of general relativity, which all physical interactions have to satisfy in order a metric description of gravity is viable. 
These conditions can be elucidated following the hypothesis introduced by Einstein \cite{Einstein:1907}, which posits strict equivalence with respect to physical laws between a coordinate system subject to a constant acceleration and a stationary one in a homogeneous gravitational field. 
Requiring the equivalence to hold only for the laws of non-relativistic physics still retains the description of gravity as a force, but already leads to the universal acceleration of free-fall. (While universality of free-fall has been known as an empirical fact at least since the VI$^{th}$ century \cite{Philoponus:500}, it remained a ``neglected clue'' prior to Einstein's work.) Extending the validity of the equivalence hypothesis to all laws of physics allows to fully equate gravitational and fictitious forces, as they cannot even in principle be distinguished. This 
further identifies inertial reference frames as the free-falling ones. Free-fall can thus be understood as an inertial motion, along a ``straight line'', albeit in a space-time that, in general, is not flat. 

Applied to special relativity, the equivalence hypothesis establishes that the space-time is a Lorentzian manifold. Requirements of the validity of the equivalence hypothesis and of special relativity together comprise the Einstein Equivalence Principle (EEP). In a modern formulation EEP is organised into three conditions \cite{clifford}: 1) Equivalence between the system's inertia and weight --  the Weak Equivalence Principle, (WEP); 2)  Independence of outcomes of local non-gravitational experiments of the velocity of a freely-falling reference frame in which they are performed (or: validity of special relativity) -- Local Lorentz Invariance (LLI);  3) Independence of outcomes of local non-gravitational experiments of their location -- Local Position Invariance (LPI).  

From the perspective of the dynamical formulation of  physical theories, the role of the EEP is to constrain the allowed form of the dynamics -- such that coordinates established by physical systems used as rods and clocks give rise to the Lorentzian space-time manifold, and the action of a system can be expressed as the length of a curve of that manifold. (Trajectories of free particles -- given by least action principle -- are then the geodesics of the manifold -- minimising the length functional of a curve.)  
The role of the EEP is thus to establish that mass-energy of a system is a universal physical quantity: inertia and weight  have to be equal in order that universality of free fall holds in the non-relativistic limit;  validity of special relativity itself requires that internal energy contributes equally to the rest mass and to inertia; and for gravitational phenomena to be equivalent to those in non-inertial frames, internal energy must contribute equally to inertia and to weight. For these reason, current tests of the EEP focus on probing the equivalence between the inertial and gravitational masses, as well as contributions of the binding energies to the mass,  for particles of different composition. 
 
This work analyses the EEP in quantum theory.  From the perspective of quantum physics classical tests involve only systems in the eigenstates of the internal energy, which is described in quantum theory as an operator. The state space of a quantum system, however, contains also arbitrary superpositions of the  internal energy eigenstates. Testing the principle for the eigenstates alone constrains only the diagonal elements of the internal energy operators, whereas to conclude about the validity of the EEP in quantum mechanics it is necessary to constrain the off-diagonal elements as well. 
We introduce a suitable quantum formulation of the EEP and the  corresponding test theory, necessary to discuss the EEP for systems with quantised internal energy. 
In order to verify the quantum formulation of the principle more parameters have to be constrained than in the classical case,  and it also requires conceptually new experimental approach. 

There is a growing interest in experiments testing the EEP with quantum metrology techniques \cite{Haensch:2004, Geiger:2011, Herrmann:2012, Schlippert:2014} as they enable using smaller test masses  and probing shorter distance scales than classical techniques \cite{Schlamminger:2008}. 
Motivation for such experiments is a general belief, that the metric picture of gravity is violated at some scale due to quantum gravity effects  \cite{Camelia:1998, Maartens:2010, Damour:2012}. In thus far realised quantum tests the mass-energies of the involved systems were still compatible with classical description. These tests still probed the equivalence of inertial and gravitational mass-energy \textit{values}, albeit in combination with the superposition principle for the centre of mass, which already merits their realisation (independently of the fully classical experiments, where also the centre of mass does not require quantum description). However, such experiments do not suffice to probe the validity of the EEP in quantum mechanics -- tests sensitive to the off-diagonal elements, e.g.\ involving superpositions of internal states,  are also required to probe it. 
 Results of this study might thus be relevant for experiments aimed at testing fundamental physics in space, for which long-term plans are currently being developed by international collaborations \cite{Kaltenbaek:2012, Aguilera:2014}. Moreover, this work provides an entirely independent motivation for quantum experiments probing the equivalence hypothesis.

\section{Massive Particle with Quantised Internal Energy}
\label{sec:massive_ham}
Hamiltonian of a massive system with quantised internal energy in the weak-field limit of Schwarzschild space-time can be found in ref.\ \cite{Zych:2011} and a general derivation valid in any static, symmetric space-time -- in ref.\ \cite{Pikovski:2013}. (As a special case of such a description one obtains relativistic Hamiltonian of a structureless massive particle \cite{ref:Laemmerzahl1996}.) In this work we only consider the lowest order relativistic corrections to the internal dynamics, since such a regime already incorporates conceptual as well as quantitative components relevant for the quantum formation of the equivalence hypothesis. Below we show how these corrections already follow from the mass-energy equivalence extended to quantum theory. 

Hamiltonian of a non-relativistic quantum system with mass $m$ subject to a gravitational potential $\phi$ reads: $\hat H_{nr}=mc^2+\frac{\hat P^2}{2m}+m\phi(\hat Q)$, where $\hat Q,\, \hat P$ are centre of mass position and momentum operators (and $mc^2$ is included just for convenience of the following arguments). Mass-energy equivalence derived from special relativity entails that increasing body's internal energy by $E$ increases also its mass by $E/c^2$ \cite{Einstein:1905}. As a result, dynamics of the system with additional internal energy $E$ is described by a Hamiltonian as above but with $m\rightarrow M:=m+E/c^2$ (currently verified up the precision of $10^{-7}$ \cite{Rainville:2005}). Note, that the mass-energy equivalence holds for any internal energy state, both in classical and quantum theory. However, in quantum theory one requires the equivalence to hold  also for arbitrary superpositions of different internal energies, due to the linear structure of the state-space of the theory. This leads to a quantum formulation of the mass-energy equivalence principle:
\begin{equation}
\label{meep}
\hat M = m\hat I_{int}+\frac{\hat H_{int}}{c^2},
\end{equation}
where $I_{int}$ is the identity operator on the space of internal degrees of freedom, $\hat H_{int}$ is the internal Hamiltonian of the system and the rest mass $mc^2$ can be defined as the ground state of the total mass-energy (i.e\ the lowest eigenvalue of $\hat H_{int}$ is zero). The dynamics of the system is described again by the Hamiltonian as above but with the mass-energy operator instead of the mass-energy parameter: $m\rightarrow \hat M$ and thus  $\hat H_{nr}\rightarrow\hat H=\hat Mc^2+\frac{\hat P^2}{2\hat M}+\hat M\phi(\hat Q)$. Hamiltonian $\hat H$ is valid up to first order corrections in $\hat H_{int}/mc^2$, and can be expanded as 
\begin{equation}
\label{meep_ham}
\hat H = mc^2+\hat H_{int}+ \frac{\hat P^2}{2m}+m\phi(\hat Q)- \hat H_{int}\frac{\hat P^2}{2m^2c^2}+\hat H_{int}\frac{\phi(\hat Q)}{c^2}.
 \end{equation}
Hamiltonian \eqref{meep_ham} is an effective description of a low-energy massive system with quantised internal dynamics, and subject to weak gravitational field. It describes the system from the laboratory reference frame. 

Mass-energy equivalence introduces the lowest order relativistic effects, described by the interaction terms: $\hat H_{int}\frac{\hat P^2}{2m^2c^2}$ and $\hat H_{int}\frac{\phi(\hat Q)}{c^2}$. The first comes from considering the inertia and the second -- from considering the weight  of the quantised internal energy. 
Since internal energy gives the rate of the internal evolution the first term describes special relativistic time dilation of the internal dynamics, 
and the second one -- gravitational time dilation \cite{Zych:2011, Pikovski:2013}. This is in full analogy to classical physics, where considering inertia and weigh of internal energy also leads to lowest order time dilation effects  \cite{Einstein:1911}. In quantum mechanics these interactions result in entanglement between internal and external degrees of freedom which results in new phenomena in quantum interference experiments with massive \cite{Zych:2011} and massless \cite{Zych:2012} systems and gives rise to a time-dilation induced decoherence \cite{Pikovski:2013}.

Note, that the Hamiltonian $\hat H=\hat Mc^2+\frac{\hat P^2}{2\hat M}+\hat M\phi(\hat Q)$ and the non-relativistic one, $H_{nr}$ have the same general structure. One might think that they thus admit the same symmetry group  
 and wonder about the nature of  relativistic effects predicted from the Eq.\ \eqref{meep_ham}. The symmetry group of $H_{nr}$ is the central extension of the Galilei group with central charge given by the mass parameter $m$ \cite{Inoenue:1952, Bargmann:1954, Levy-Leblond:1963}, whereas the symmetry group of $\hat H$ has central charge given by $\hat M$ -- an operator which acts on a Hilbert space of the internal degrees of freedom (such as vibrational or electromagnetic energy levels of an atom). Note, that a superposition of eigenstates of $\hat H_{int}$  -- and thus of $\hat M$ -- evolves in time and will  exhibit time dilation effects explained in the previous paragraph. Thus, non-trivial central extensions of the Galilei group do not describe a non-relativistic theory, the latter  shall give rise to an absolute time of Euclidean space-time. A consistent non-relativistic limit for a system with internal dynamics is therefore given not only by restricting to small centre of mass energies, but also to slow internal evolution, such that dynamical contributions to the mass-energy operator are small compared to the static contribution.  In such a case, the dynamical part of the mass-energy only contributes to the rest mass-energy $Mc^2$  (which allows for a correct description of non-relativistic systems with internal degrees of freedom, where  internal energy adds up to the energy of the centre of mass) but only the static part effectively contributes to the mass-energy in the kinetic and potential energy terms. Formally, this is tantamount to requiring that internal energy becomes effectively fully degenerate in the non-relativistic limit, with all states having the same value of the mass-energy. Internal energy eigenstates are stationary and will not exhibit any relativistic effects.  Note, that such an operational way of defining the non-relativistic limit -- as the regime where effectively all relativistic  effects, including the time dilation of internal evolution, are suppressed --   can be seen as the origin of the split between mass and energy which are fully equivalent in relativity (and allows \textit{defining} the mass as the static part of the internal energy). Such an approach also sheds a very different light on the Bargmann's superselection rule for the mass \cite{Bargmann:1954}, which we further discuss in the Appendix \ref{app:conditions}. Finally, note that the mass-energy equivalence provides a natural physical interpretation of non-trivial central charge operators of  central extensions of the Galilei group -- in terms of an internal energy of a composite system, which contributes to the mass and generates evolution of the internal degrees of freedom.


\section{The model}
\label{sec:model}
We now construct a test model for analysing the validity of the EEP in quantum theory, which reduces to Eq.\ \eqref{meep_ham} if the principle is valid.  
We generalise a standard approach to constructing test theories, in which possibly different inertial and gravitational mass parameters $m_i$ and $m_g$ and internal energy values are considered. Additionally,  we allow that the entire mass-energy operators can have distinct gravitational $\hat M_{g}$ and inertial $\hat M_{i}$ form, and that both can differ from the rest mass-energy operator, $\hat M_{r}$. We thus introduce a modified quantum formulation of the mass-energy equivalence Eq.\ \eqref{meep}:
\begin{equation}
\label{masses}
\hat M_\alpha:= m_\alpha\hat I_{int}+\frac{\hat H_{int, \alpha}}{c^2}\; \alpha=r,i,g,
\end{equation}
where $\hat H_{int, r}$ is the rest energy operator (operationally defined as the Hamiltonian  in the rest frame of the system, in a region far away from massive objects), $\hat H_{int,i}$ and $\hat H_{int, g}$ are the contributions of the internal energy to  $m_i$ and $m_g$, respectively. Rest mass parameter $m_r$ is not observable in the present context and can be assigned an arbitrary value without changing predictions of the model (it acquires physical meaning of active gravitational mass when gravitational field generated by the system is considered). With the mass-energy operators of Eq.\ \eqref{masses} we obtain the following test Hamiltonian: $\hat H_{test}=\hat M_r+\frac{P^2}{2\hat M_i}+\hat M_g\phi(Q)$ which is valid to the lowest order in relativistic corrections:
\begin{equation}
\label{ham}
\hat H^Q_{test}= m_rc^2+\hat H_{int,r}+\frac{\hat P^2}{2m_i}+m_g\phi(\hat Q) - \hat H_{int,i}\frac{\hat P^2}{2m_i^2c^2}+\hat H_{int,g}\frac{\phi(\hat Q)}{c^2}.
\end{equation}
Hamiltonian $\hat H^Q_{test}$ constitutes a new model for analysing the the EEP in a regime where the relevant degrees of freedom  -- internal mass-energies -- are quantised. It incorporates non-trivial expression of all the three conditions into which the EEP is organised. Validity of the WEP requires equivalence between inertia and weight and its quantitative expression in our model reads $\hat M_{i}=\hat M_{g}$. As all the relativistic effects in the considered regime are  derived from the mass-energy equivalence, the validity of special relativity, LLI, is to lowest order expressed by requiring that internal energy contributes equally to the rest mass and to inertia: $\hat H_{int,r}=\hat H_{int, i}$, analogously,  LPI is expressed by requiring that rest energy equally contributes to weight: $\hat H_{int,r}=\hat H_{int, g}$. Keeping in mind that $m_r$ can be assigned arbitrary value without changing the physics of the model, the validity of the EEP (to lowest order) is in quantum theory expressed by $\hat M_{r}=\hat M_{i}=\hat M_g$. For $n$-level quantum system testing validity of the EEP thus requires measuring $2n^2-1$ real parameters (comparing elements of hermitian operators $M_\alpha$, where one parameter, $m_r$, is free).  

In Appendix \ref{app:conditions} we re-derive $\hat H^Q_{test}$ and the conditions for the validity of the EEP directly from imposing validity of the Einstein's hypothesis of equivalence on the dynamics of a low-energy relativistic quantum system with internal degrees of freedom, showing that the two approaches are indeed equivalent.

In a model with  internal energy incorporated as classical parameters the conditions expressing validity of the EEP are just a special case of the quantum conditions derived above. See Table \ref{table:violations} for a summary. Such model can be described by
\begin{equation}
H^C_{test}= M_r+\frac{\hat P^2}{2M_i}+M_g\phi (\hat Q) \approx m_rc^2+E_r+\frac{\hat P^2}{2m_i}+m_g\phi (\hat Q)-E_i\frac{\hat P^2}{2m_ic^2}+E_g\frac{\phi(\hat Q)}{c^2}.
\end{equation} 
where $M_\alpha:=m_\alpha+E_\alpha/c^2$ denote the total mass-energies with $E_r$ the value of internal energy contributing to the rest mass; $E_i$ -- to the inertial mass, and $E_g$ -- to the weight. In the test model $H^C_{test}$  WEP is expressed  by requiring $M_i=M_g$, LLI -- by requiring $E_r=E_i$ and LPI -- by $E_r=E_g$, for each internal state.  These conditions are the same for a fully classical theory (see also Appendix \ref{app:classical}). Although $H^C_{test}$ incorporates quantised centre of mass degrees of freedom it incorporates only the classical formulation of the EEP, from this perspective is equivalent to a classical test theory.  The classical conditions above can be seen as a restriction of the quantum requirements to the diagonal elements of the internal energy operators (or only to the operators' eigenvalues). The quantum conditions reduce to the classical ones only if an additional assumption is made -- that operators $ \hat H_{int, \alpha}$ mutually commute. For a system with $n$ internal classical states testing validity of the EEP requires measuring only $2n-1$ parameters. Therefore, validity of the EEP in quantum physics is not guaranteed by its validity in classical theory and requires conducting independent experiments. (Only in the non-relativistic limit quantum and classical formulations of the  EEP coincide, as both reduce to the requirement $m_g=m_i$,  which is  the non-relativistic expression of the  WEP.)
\renewcommand{\arraystretch}{1.6}
\begin{table}[h!]
\caption{Conditions for the validity of the EEP in the classical and in the quantum theory and number of parameters to test it (to lowest order) for a system with $n$ internal states. In the non-relativistic limit the EEP reduces to the Weak Equivalence Principle (WEP), and only requires equivalence of the inertial $m_i$ and the gravitational $m_g$ mass parameters. Validity of the Local Lorentz Invariance (LLI) and of the Local Position Invariance (LPI) guarantees universality of special and general relativistic time dilation of the internal dynamics, respectively. In quantum mechanics their validity requires equivalence of rest, inertial and gravitational internal energy \textit{operators} $\hat H_{\alpha}$, $\alpha=r,i,g$. In the classical case, it suffices that the \textit{values} $E_\alpha$ of the corresponding internal energies are equal. Quantities $E_\alpha$ can be seen as the diagonal elements of $\hat H_{int,\alpha}$ and thus, beyond the Newtonian limit,  validity of the EEP in classical mechanics does not guarantee its validity in quantum theory. 
\hspace*{\fill}}
\centering
\begin{tabular}{|c c|c| c| c| c|}
   \multicolumn{1}{c}{}    &   \multicolumn{1}{c}{}   &      \multicolumn{1}{c}{}      &  \multicolumn{1}{c}{{\bf EEP}}    &    \multicolumn{1}{c}{}  & \multicolumn{1}{c}{}  \\ \cline{3-6}
  \multicolumn{1}{c}{}   &                                    & \textbf{WEP}                       &   \textbf{LLI}        & \textbf{LPI}        & \textbf{\# param.} \\ \hline

{Newtonian}& classical \& quantum             & $m_i= m_g$                          &   $-$                  &  $-$                  &      1                   \\ \hline
Newtonian +& classical 					 & $m_ic^2+{E_i}= m_gc^2+{E_g}$ & $E_r =E_i$    & $E_r = E_g$      &       $2n-1$          \\ \cline{2-6}
mass-energy equiv.& quantum  &  $m_ic^2 \hat I+\hat H_i= m_g c^2 \hat I+\hat H_g$ & $\hat H_r = \hat H_i$ & $\hat H_r = \hat H_g$ &    $2n^2-1$ \\ \hline
\end{tabular}
\label{table:violations}
\end{table}

Modern quantum tests of the EEP are performed with composed systems, like atoms, in interferometric experiments where internal energy levels are used to manipulate external degrees of freedom of the system \cite{Haensch:2004, Geiger:2011, Herrmann:2012, Schlippert:2014}. Currently, in analysis of these tests internal atomic energy is treated classically. This is consistent only as long as these experiments remain probing  the non-relativistic limit of the EEP, the universality of free fall of  Newtonian gravity. The assumption that internal energy can be treated classically will inevitably be violated when the experimental precision increases. Moreover, in such a regime tests involving only internal eigenstates will not be sufficient to verify the validity of the EEP. Experiments with systems in superpositions of internal energy states will be required and for a meaningful analysis of such experiments a
test theory incorporating quantised internal energy, like $\hat H^Q_{test}$, will be necessary.

Quantised internal energy has not been previously incorporated into theoretical frameworks for analysing the EEP in quantum mechanics. Models studied thus far introduce a modified Lagrangian (or action) like e.g.: the\ $TH\epsilon\mu$-formalism \cite{Lee:1973}, Standard Model Extensions (SME) \cite{Kostelecky:1997} or a modified Pauli equation \cite{Laemmerzahl:1998}. Possibility of violations of the EEP is incorporated by introducing distinct inertial and gravitational mass parameters, (spatial) mass-tensors \cite{Haugan:1979} or spin-coupled masses \cite{Laemmerzahl:1998} for elementary particles or fields. In these approaches, for describing bound systems one derives from the elementary model an effective one, with EEP-violating parameters which describe shifts in values of the binding energies, see e.g.\ \cite{Hohensee:2013},  but the dynamics of the degrees of freedom associated to the binding energy has not been considered. Note however, that if the fundamental interactions are modified, not only the eigenvalues but also the eigenstates of the effective internal Hamiltonians of composed systems will be different. This shows that here discussed features of the quantum formulation of the EEP are generic in theories incorporating EEP violations at the level of fundamental interactions, which are themselves  anticipated in most studies of quantum gravity phenomenology \cite{Camelia:1998, Maartens:2010, Damour:2012}.  
For a direct comparison between our Hamiltonian approach and the Lagrangian-based frameworks we derive the Lagrangian formulation of our test theory in the Appendix \ref{app:lagrangian}.

Internal degrees of freedom were thus far considered only in the context of the WEP for neutrinos \cite{Gasperini:1988, Mann:1996}. It has been studied how neutrino oscillations would be affected if the neutrinos' weak interaction eigenstates would be different from the states with a well-defined value of the gravitational mass. (For massless neutrinos such effects are excluded with a precision of $10^{-11}$ \cite{Gasperini:1988} and for models of massive neutrinos are ruled out by the existing experimental data \cite{Mann:1996}.)


Understanding the (in)dependence relations between the three tenets of the EEP is important not only for counting the number of parameters to test. For the field of precision tests of the EEP particularly important is the question: under what assumptions tests of the WEP impose constraints on the violations of LPI? First, note that no single principle implies the others -- see e.g.\ Table \ref{table:violations}. Second, test theory $\hat H^Q_{test}$ is based on three assumptions: 1) energy is conserved, 2) in the non-relativistic limit standard quantum theory is recovered when inertial and gravitational mass parameters are equal, 3)  mass-energy equivalence extends to quantum mechanics as in Eq.\ \eqref{masses}. With these assumptions alone validity of the WEP ($\hat M_g=\hat M_i$)  does not imply validity of LPI ($\hat H_{int,r}=\hat H_{int,g}$). Additional assumption has to be made: 4) LLI is valid ($\hat H_{int,r}=\hat H_{int,i}$). Only under all four assumptions  constraints on the violations of the WEP can in principle give some constraints on the violations of the LPI. Note, that all the four assumptions are in fact made also in the well-known Nordtvedt's gedanken experiment \cite{Nordtvedt:1975}, which is often invoked as a proof that energy conservation alone yields tests of the WEP to be equivalent to tests of LPI. We stress that the three tenets of the EEP concern very different aspects of the theory, which itself merits their independent experimental verification.

\section{Testing the quantum formulation of the EEP}
\label{new_effects}
Approach developed in this work allows introducing an unambiguous distinction between tests of the classical and of the quantum formulation of the EEP: As tests of the classical formulation of the  EEP qualify experiments whose potentially non-null results (indicating violations of with the EEP) can still be explained by a diagonal test theory -- i.e.\ test theory where internal energy operators are assumed to commute. (E.g.\ by a model like $H^C_{test}$ which can be seen as a diagonal element of $\hat H^Q_{test}$ in a special case when all  $\hat H_{int,\alpha}$ commute). In contrast, experiments which non-null results cannot be explained by a diagonal model, can be seen as tests of the quantum formulation of the EEP. Below we discuss some experimental scenarios for testing various parts of the EEP and show which aspects of such experiments test the quantum and which test the classical formulation of the EEP. 

In order to put quantitative bounds on possible violations of the quantum formulation of the EEP it is convenient to introduce a suitable parametrisation: Violations of the quantum formulation of the WEP will be described by a parameter-matrix $\hat \eta:=\hat I_{int}-{\hat M_g}{\hat M_i}^{-1}$; of the LLI by $\hat \beta:=\hat I_{int}-\hat H_{int,i}\hat H_{int,r}^{-1}$; and of the LPI by $\hat \alpha:=\hat I_{int}-\hat H_{int,g}\hat H_{int,r}^{-1}$. In order to parametrise violations of the classical formulation of the EEP we need: for the WEP -- a real parameter $\eta_{class}:=1-M_g/M_i$ for each internal state; for  LLI and LPI we need one parameter $\beta_{class}:=1-E_i/E_r$ and one $\alpha_{class}:=1-E_g/E_r$, for each internal state (apart from, say, the ground state which can be set arbitrarily through a free parameter $m_r$). The classical parameters $\eta_{class}, \alpha_{class}, \beta_{class}$ can again be seen as the diagonal elements of the quantum parameter-matrices: $\hat \eta$, $\hat \alpha$, $\hat \beta$. Note, that there is a total of $2n^2-1$ independent real parameters for testing the quantum and $2n-1$ for the classical formulation of the EEP as explained in Sec.\ \ref{sec:model}.

\subsection{Testing the quantum formulation of the WEP}
The physical meaning of the validity of the WEP in classical theory is the universality of free fall -- which is also how WEP is usually tested. 
In quantum theory, universality of free fall can be generalised to the requirement that the ``acceleration'' the position operator for the centre of mass is independent of the internal degrees of freedom. This is best seen in the Heisenberg picture, where time evolution of an observable $\hat A$ under a Hamiltonian $\hat H$ is given by  $d\hat A/dt=-i/{\hbar}[\hat A,\hat H]$. 
Under $\hat H^Q_{test}$ the acceleration of the centre of mass $\hat{a}_{\hat H^Q_{test}}:=d^2\hat Q/dt^2=-\frac{1}{\hbar^2}[[\hat Q, \hat H^Q_{test}], \hat H^Q_{test}]$ is given by
\begin{equation}
\label{q_uff}
\hat{a}_{\hat H^Q_{test}}=-{\hat M_g}{\hat M_i}^{-1}\nabla \!\phi(\hat Q)+\frac{i}{\hbar}[\hat H_{int,i}, \hat H_{int,r}] \frac{\hat P}{m_ic^2} +\mathcal{O}(1/c^4).
\end{equation}   
Let us first note that the commutator term in Eq.\ \eqref{q_uff} is present already in vanishing gravitational potential $\phi(Q)$ and expresses a violation of LLI. 
Under the Hamiltonian $H^C_{test}$ the centre of mass acceleration $\hat{a}_{H^C_{test}}$ reads
\begin{equation}
\label{tilde_uff}
\hat{a}_{H^C_{test}}=-{M_g}{M_i}^{-1}\nabla \!\phi(\hat Q)\,
\end{equation}   
(as expected, it can be obtained from Eq.\ \eqref{q_uff} as a special case of commuting internal energy operators). From Eqs \eqref{q_uff} and \eqref{tilde_uff} follows that also in quantum theory probing the WEP is tantamount  to probing whether the time evolution of the centre of mass degree of freedom is universal, does not depend on the state of the system.  

\begin{figure}[h]
\centering
\includegraphics[trim = 3.4cm 0mm 0mm 0mm, clip=true, width=11cm]{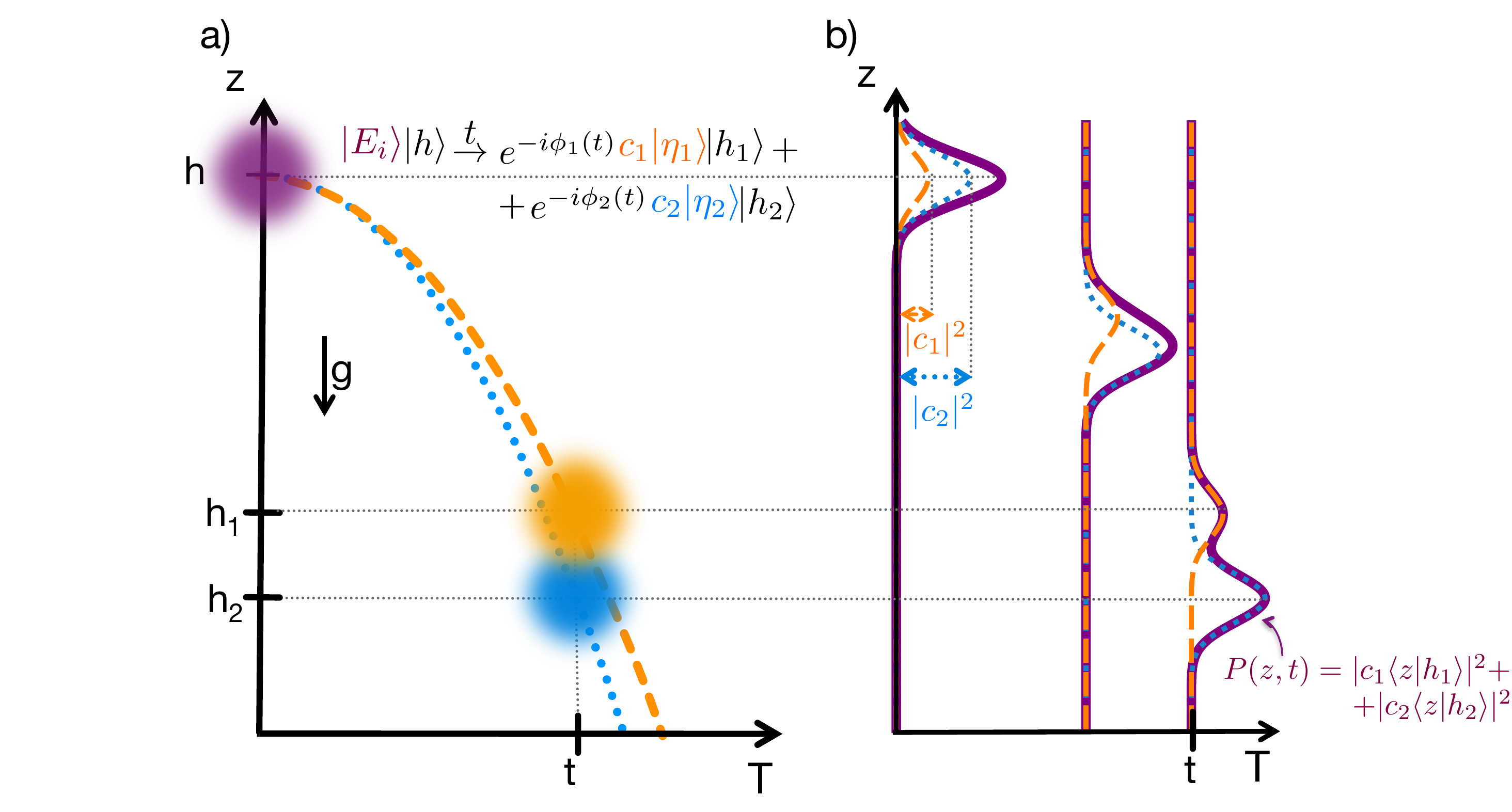}
\caption{Free evolution of the centre of mass (c.m.) degree of freedom (d.o.f.) of a quantum system in a gravitational field $g$ in a presence of the Weak Equivalence Principle (WEP) violations. Initially state of the system is a product of an internal state $|E_i\ket$ and c.m.\ position $|h\ket$ given by a gaussian distribution centred at height $h$. If the quantum formulation of the WEP is violated  the system is in superposition of c.m.\ states each falling with different accelerations. \textbf{a)} Dashed orange and dotted blue lines represent semi-classical trajectories of the c.m.\ correlated with the internal states $|\eta_{1}\ket, |\eta_2\ket$, for which acceleration is well defined $\eta_1g, \eta_2g$, marked in the corresponding colours. \textbf{b)} Probability distribution $P(z,t)$ of finding the system at height $z$ at time $t$ (see main text) is marked by a purple line. Dashed orange and dotted blue lines represent the probability conditioned on the internal state $|\eta_{1}\ket$ and $|\eta_2\ket$, respectively. Modulations in $P(z,t)$ indicate violation of the quantum formulation of the WEP under the assumption that linearity of quantum theory is not violated. Probing entanglement  between internal and c.m.\ d.o.f. generated in the above scenario, would constitute a direct test of the quantum formulation of the WEP.\hspace*{\fill}}
\label{QUFF}
\end{figure}

Classical WEP violations,  effects derivable from $\hat M_i \neq \hat M_g $ but $[\hat H_{int,i}, \hat H_{int,g}]= 0$, would result in different accelerations for different internal states. Violations stemming from $[\hat H_{int,i}, \hat H_{int,g}]\neq 0$ result in additional effects. Assume that LLI holds. Eq.\ \eqref{q_uff} and a relation  ${\hat M_g}{\hat M_i}^{-1}=\hat I_{int}-\hat \eta$ entail that only eigenstates of the parameter-matrix $\hat\eta$, which  explicitly reads $\hat \eta\approx m_g/m_i(\hat I+\hat H_{int,g}/m_gc^2-\hat H_{int, i}/m_ic^2)$, have a well defined free fall acceleration of the centre of mass degree of freedom. Hence, for a given internal energy eigenstate the external degree of freedom will in general be in a superposition of states each falling with a different acceleration, see Fig.\ \ref{QUFF} a).  Consider an eigenstate of $\hat H_{int,i}$ initially semi-classically localised at some height $|h\ket$: $|\Phi(0)\ket=|E_i\ket|h\ket$. (An eigenstate of $\hat H_{int,i}$ can be prepared e.g.\ with a precise mass-spectroscope, as it allows selecting states of a given inertial mass-energy.) Under free fall it would generally evolve into $|\Phi(t)\ket=\Sigma_i e^{-i\phi_i(t)}c_i |\eta_i\ket|h_i\ket$, where $|\eta_i\ket$ are eigenstates of $\hat \eta$, $c_i$ are normalised amplitudes such that: $|E_i\ket=\Sigma_i c_i |\eta_i\ket$, and $\phi_i(t)$ are the phases acquired during the evolution. When $|h_i-h_j|$ for different $i,j$ become larger than the system's coherence length,  the position amplitudes become distinguishable, $\bra h_j|h_i\ket=\delta_{ij}$, and the state $|\Phi(t)\ket$ becomes entangled: with the centre of mass position entangled to the internal degree of freedom. Probing this entanglement would constitute a direct test of the quantum formulation of the WEP, since under a ``diagonal test model'' (model in which internal energy operators commute) initially separable state $|\Phi(0)\ket$ cannot evolve into an entangled one.

As a result of the above described entanglement, the probability of finding the system at time $t$ as a function of the  height $z$, $P(z,t)=|\bra z|\Phi(t)\ket|^2$, develops distinguished spatial modulations, see Fig.\ \ref{QUFF} b). In the opposite limit,  when the coherence length dominates over $|h_i-h_j|$,  the spatial probability distribution just broadens in the direction of gravity gradient. For an initial eigenstate of $\hat H_{int, i}$ such modulations or broadening would not appear if $[\hat H_{int,i}, \hat H_{int,g}]=0$, unless it allows that a pure state $|\Phi(0)\ket$ evolves into a mixed one $\hat \rho(t):=\Sigma_i |c_i|^2 |\eta_i\ket|\bra \eta_i|\otimes|h_i\ket\bra h_i|$. 
Such a model does not violate the quantum formulation of the WEP,  but it violates unitarity of quantum theory. It can explain the broadening or the spatial modulations of the probability distribution,  since $|\bra z| \hat \rho(t)|z\ket|^2\equiv P(z,t)$, but cannot account for the entanglement as $\hat\rho(t)$ is separable. Thus, probing such broadening or modulations can be considered a test of the quantum WEP under an additional assumption that linear structure of quantum theory is retained. Note however, that such additional broadening would be very difficult to measure precisely and to distinguish from the standard quantum mechanical effects causing spreading of the particles' wave-packets.

A recent experiment realised in a drop tower in Bremen with Bose Einstein Condensate (BEC) of $^{87}$Rb in extended free fall \cite{Muentinga:2013} can be used to put some bounds on the strength of such modulations and  constrain some of the new parameters of $\hat H^Q_{test}$.
A small, but non-zero variance of the parameter-matrix $\hat\eta$, denote by $\Delta \eta$, would lead to an anomalous spreading of the free falling  BEC cloud by $\Delta S\approx \Delta\eta gT^2/2$, where $T\approx 0.5$s denotes the free-fall time and $g\approx 10$m$/s^2$. As no anomalous spreading or modulations of the BEC cloud has been reported we assume that  $\Delta S$ can be bounded by the size of the BEC cloud, which we estimate to be $L\approx 10^{-4}$m; as a result $\Delta\hat \eta<8\cdot10^{-5}$. Under the assumption that the initial state of the atoms was an eigenstate of $\hat H_{int,i}$ and that unitarity of quantum theory is not violated, a non-zero $\Delta \hat \eta$ could only arise from $[\hat H_{int,i}, \hat H_{int,g}]\neq 0$. 


 \subsection{Testing the quantum formulation of the LPI and LLI }
Validity of LLI and LPI can be tested in experiments probing special relativistic and the gravitational time dilation, respectively: Allowing for different internal Hamiltonians $\hat H_{int,\alpha}$ in general results in a  different speed of the internal evolution. Denoting the internal degree of freedom by $\hat q$ the Hamiltonian  $\hat H^Q_{test}$ yields
\begin{equation}
\label{q_time_dilation}
\dot {\hat q}(\hat Q,\hat P) = \dot {\hat q}_r\hat I_{ext}-\dot {\hat q}_i  \frac{\hat P^2}{2m_i^2c^2}+\dot {\hat q}_g\frac{\phi(\hat Q)}{c^2},
\end{equation}
where  $\dot{\hat q}_\alpha:=-i/\hbar[\hat q, \hat H_{int,\alpha}]$ and $\hat I_{ext}$ is the identity operator on the space of external degrees of freedom $(\hat Q, \hat P)$. In terms of the velocity
$\dot Q$ canonically conjugate to the momentum $P$ (see also Appendix C) we can write Eq.\ \eqref{q_time_dilation} in the form $\dot {\hat q}(\hat Q,\dot{\hat Q}) = \dot {\hat q}_r\hat I_{ext}-\dot {\hat q}_i  \frac{\dot{\hat Q}}{2c^2}+\dot {\hat q}_g\frac{\phi(\hat Q)}{c^2}$.
If internal energy is coupled universally, $\hat H_{int, \alpha}=\hat H_{int}$ for $\alpha=i,r,g$, we have $\dot{\hat q}_{\alpha}=\dot{\hat q}$ and thus: $\dot {\hat q}(\hat Q,\hat P) = \dot {\hat q}(\hat I_{ext}-\frac{\hat P^2}{2m_i^2c^2}+\frac{\phi(\hat Q)}{c^2})$ -- i.e.\ we recover universal special relativistic and gravitational time dilation of the internal dynamics (up to lowest order in $c^{-2}$). Mass parameters $m_\alpha$ are irrelevant for the rate of the internal  dynamics and thus in the non-relativistic limit there is no time dilation, the internal evolution is just given by the rest energy operator. The condition for the gravitational time dilation to be universal reads $\hat H_{int,g}=\hat H_{int,r}$ and analogously for the special relativistic time dilation: $\hat H_{int,i}=\hat H_{int,r}$. Testing universality of the time dilation effects is therefore equivalent to probing LPI  and LLI (to lowest order), see also Appendix \ref{app:conditions}. 

In any test-theory incorporating classical internal energy  one can at most consider  special relativistic and gravitational redshifts of the internal energy. Analogously to the above discussed case, in the theory described by $H^C_{test}$ special relativistic redshift is universal once $E_r=E_i$ and the gravitational redshift is universal if $E_r=E_g$. These conditions also hold in a fully classical theory, see the also Appendix \ref{app:classical}. In turn, this entails that an experiment measuring only the redshift of atomic spectra or only the time dilation of clocks following classical paths, can always be explained via LLI  or LPI violations which are  compatible with $[\hat H_{int,r}, \hat H_{int,i}]=0$ or $[\hat H_{int,r}, \hat H_{int,g}]=0$, respectively. Without additional assumptions or measurements of additional effects, such experiments can only be seen as tests of the classical formulation of the equivalence principle.

\begin{figure}[h!]
\centering
\includegraphics[ width=13cm]{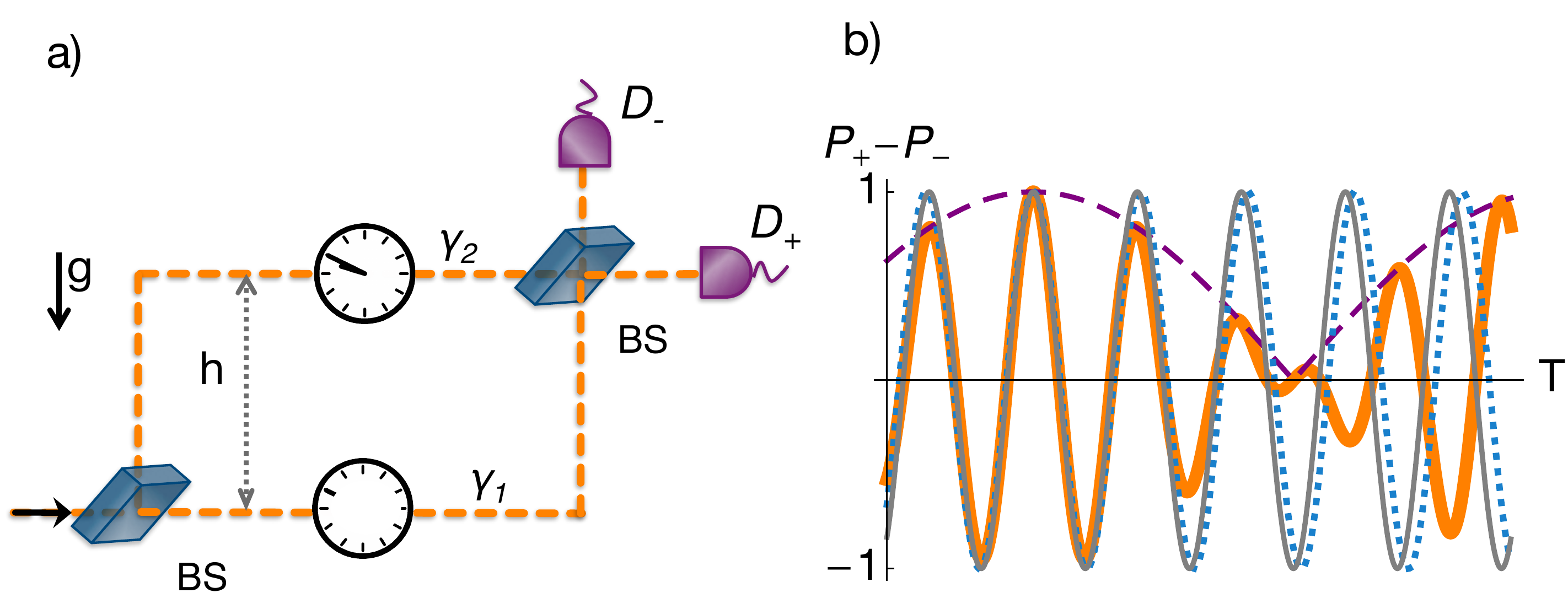}
\caption{ \textbf{a)} Mach-Zehnder interferometer for probing the quantum formulation of the Local Position Invariance (LPI) and \textbf{b)} detection probabilities in different scenarios. The setup consists of two beam splitters (BS) and two detectors $D_\pm$, is stationary in the laboratory reference frame subject to gravitational field $g$. The setup permits  two fixed trajectories $\gamma_1$, $\gamma_2$ with separation $h$ in the direction of the field. Initial internal state of the system is stationary along $\gamma_1$. If LPI is respected, gravitationally induced phase shift of the interference pattern will be observed --  thin, grey line --  with in principle maximal visibility.  Observing a different phase shift   -- dotted blue line -- indicates a violation, but it can always be explained by a diagonal test theory. Modulations of the visibility -- purple, dashed line --  cannot be explained with a diagonal model (for here chosen intial state) and arise directly from the non-commutativity of the internal energy operators. Thick, orange line represents detection probabilities in this case. Internal d.o.fs stationary along $\gamma_1$ will generally not be stationary along $\gamma_2$ and the system is therefore represented as a ``good clock'' only along $\gamma_2$. The resulting entanglement between internal and external d.o.fs leads to the modulation in the fringe contrast. Measurements of the visibility of the gravitationally induced interference pattern can thus probe the quantum formulation of the LPI, whereas measurements of the phase shift alone can always be seen as probing only the classical formulation.\hspace*{\fill}}
\label{QLPI}
\end{figure}

Violations of the quantum formulation of the LLI (LPI)  coming from $[\hat H_{int,r}, \hat H_{int,i}]\neq0$ ($[\hat H_{int,r}, \hat H_{int,g}]\neq0$)  lead to conceptually different effects, since the non-commuting operators generally have different stationary and time-evolving states -- an eigenstate of, say,  $\hat H_{int,r}$ will generally not be an an eigenstate of  $\hat H_{int,i}$ ($\hat H_{int,g}$). 
Consider an interference experiment where a particle follows in superpositions two different semi-classical trajectories $\gamma_1, \gamma_2$ which are then coherently overlapped and the resulting interference pattern is observed, see Figure \ref{QLPI} a) for a sketch of the setup. If the centre of mass degrees of freedom are constrained to follow a semi-classical path $\gamma_j$, $j=1,2$ the total Hamiltonian describes the dynamics of the internal degrees of freedom along this path and we denote it by $\hat H^Q_{test}(\gamma_j)$. If the initial internal state is an eigenstate $|E(\gamma_1)\ket$ of $\hat H^Q_{test}(\gamma_1)$ it remains stationary along path $\gamma_1$ like a ``rock''. 
However, it will generally non-trivially evolve in time along $\gamma_2$, like a ``clock'',  if $[\hat H^Q_{test}(\gamma_1),\hat H^Q_{test}(\gamma_2)]\neq0$.  
As a result, the internal state of the particle entangles to the centre of mass  and the coherence of the centre of mass superposition decreases. This loss of coherence is given by the overlap between the two internal amplitudes evolving along different paths. 
For a quantitative analysis assume that gravitational potential is approximately homogeneous $\phi(x)=gx$  and paths $\gamma_j$ are defined such that the particle remains at rest in a laboratory frame at fixed height $h_j$ for time $T$; so that  $\hat H^Q_{test}(\gamma_j)=\hat H_{int,r}+(m_gc^2+\hat H_{int,g})\frac{gh_j}{c^2}$ and the contributions from the vertical parts of $\gamma_j$, see Figure \ref{QLPI}, are the same.
For such a case the coherence of the centre of mass superposition reads
 $\mathcal{V} \approx |\cos\left(\Delta H_{int,g}\frac{ghT}{\hbar c^2}\right)|$,
 where $h\equiv h_2$, $h_1=0$ and with $\Delta  H_{int,g}:=\sqrt{\bra \hat H_{int,g}^2\ket-\bra \hat H_{int,g}\ket^2 }$, where expectation values are taken with respect to the initial state $|E(\gamma_1)\ket$, i.e.\ the eigenstate of $\hat H_{int,r}$. 
Quantity $\mathcal{V}$  is also  the visibility of the interference pattern  observed in such an experiment:  the probabilities $P_\pm$ to detect the particle in the detector $D_\pm$ read in this case
 $P_\pm\approx\frac{1}{2}\left(1\pm \cos\left(\Delta H_{int,g}\frac{ghT}{\hbar c^2}\right)\cos\left( \bra \hat M_g\ket \frac{ghT}{\hbar}\right) \right)$.  The first term is the overlap between the amplitudes  that were evolving along different paths, the second comes from the relative phase-shift between them.  
In a general case of arbitrary paths $\gamma_j$ and arbitrary initial state, the detection probabilities read $P_\pm=\frac{1}{2}\left(1\pm \mathrm{Re} \{\bra e^{\frac{i}{\hbar} \int_{\gamma_1}ds\hat H^Q_{test}\left(\gamma_1(s)\right)} e^{-\frac{i}{\hbar} \int_{\gamma_2}ds\hat H^Q_{test}\left(\gamma_2(s)\right)}\ket \}\right)$ and the visibility reads $\mathcal{V}=|\bra  e^{\frac{i}{\hbar} \int_{\gamma_1}ds\hat H^Q_{test}\left(\gamma_1(s)\right)} e^{-\frac{i}{\hbar} \int_{\gamma_2}ds\hat H^Q_{test}\left(\gamma_2(s)\right)}\ket|$, where the expectation values are taken with respect to the initial state of the system.

If the EEP is valid, modulations in the visibility in such an interference experiment only occur if an initial state has a non-vanishing internal energy variance  and if there is time dilation between the paths $\gamma_j$ \cite{Zych:2011, Pikovski:2013}. An internal energy eigenstate remains an eigenstate in the standard theory, independently of the path taken by the centre of mass and only results in a phase shift term, with in principle maximal visibility $\mathcal{V}=1$. In a diagonal test theory the initial state would also remain an eigenstate along both paths, with at most different eigenvalues of the different internal  energies.  This would result  in modifications of the observed phase shift, but would allow allow for a maximal visibility. 
Modulations in the visibility of the inference pattern in the experiment with a system prepared in  internal eigenstate can thus  probe the quantum formulation of the EEP.


Let us rewrite $P_\pm$ in terms of the parameters quantifying violations of the quantum and of the classical formulation of the EEP. What aspects of the EEP are tested depends on which parameters  are independently measured and which are inferred from the test. In the considered scenario the system is prepared in an eigenstate of $\hat H_{int,r}$. One can thus rewrite  $\Delta H_{int,g}\equiv E_r\Delta \alpha$ where $\Delta\alpha$ is a variance of a parameter-matrix $\hat \alpha:=\hat I_{int}-\hat H_{int,g}\hat H_{int,r}^{-1}$. In many atom interference experiments the separation between the paths is not measured directly but is given by a momentum transfer from a laser pulse. In such a case $h=\hbar k t_s/\bra \hat M_i\ket $, where $k$ is a wave-vector of the light pulse and $t_s$ is the time after which in our scenario the amplitudes remain at fixed heights. Note, that $\bra \hat M_g\ket/\bra\hat M_i\ket \neq1$ can always be explained by commuting $\hat M_g$, $\hat M_i$ with only different eigenvalues; one can identify $\eta_{class}=1-\bra \hat M_g\ket/\bra\hat M_i\ket$. Without further assumptions
$$P_\pm=\frac{1}{2}\left(
1\pm \cos\left(\Delta \alpha\frac{E_r}{\bra M_i\ket c^2}gkt_sT\right)
\cos\left((1-\eta_{class})gkt_sT\right)\right).$$
The term proportional to $(1-\eta_{class})$ describes the gravitational phase shift and its possible modifications due to the violations of the classical formulation of the WEP. Moreover, the above entails that any modification to the phase shift in a typical interference scenario can  be fully explained by a model that only violates the Newtonian limit of the the WEP (i.e.\ a test theory with $m_i\neq m_g$ and $\hat H_{int,i}=\hat H_{int,r}=\hat H_{int,g}$).  The visibility of this interference patter, the first cosine, allows probing the quantum formulation of the LPI as it  depends on the variance of the quantum parameter-matrix $\hat \alpha$. 

%


No experiment has yet succeeded  to probe jointly general relativistic and quantum effects. No direct bound thus exist on the quantum violations of the LPI.  By assuming validity of the LLI some constraints can be inferred e.g.\ from an interferometric experiment realised in the group of T.\ H\"{a}nsch \cite{Haensch:2004} measuring gravitational phase shift for two Rubidium isotopes $^{85}\textrm{Rb}$ and $^{87}\textrm{Rb}$, and for two different hyperfine states of $^{85}\textrm{Rb}$.  For the quantitative analysis  we take $\bra\hat M_r\ket=\bra \hat M_i\ket\approx 2.5 \cdot 10^{-25}$ kg for the mass of Rubidium and $E_r=\hbar \omega$ with $\omega\approx 10^{15}$ Hz. The wave vector of the light grating used in the experiment was $k\approx7.5\cdot10^6$ m$^{-1}$  and to obtain the interference pattern the time $T$ was varied between $40.207$ and $40.209$ms.  We can set $T=T_0\pm\delta T$ with $T_0=40$ms and $\delta T =10^{-3}$ms. The height difference achieved with this interferometric technique was $h\approx\hbar k T_0/\bra M_r\ket$.  Experimentally measured visibility remains constant over the interrogation times $T_0\pm\delta T$ and was equal $\mathcal V=0.09$.  Any variations of the visibility must therefore be of order $10^{-3}$ or smaller. This experiment thus allows to obtain a bound $\Delta\alpha<9\cdot10^6$. 

For testing LLI an interferometric experiment analogous to the one in the Figure \ref{QLPI} can be used, but with a horizontal setup where the two interferometric paths remain at the same gravitational potential. One amplitude, $\gamma_1'$, is kept at rest with respect to the laboratory frame whereas the other, $\gamma_2'$, is given some velocity and after reflection is overlapped with the first one. If $[\hat H_{int,r}, \hat H_{int,i}]\neq0$ even if the internal state of the interfering system is stationary along $\gamma_1'$, it will generally not be stationary along $\gamma_2'$. In a full analogy to the above discussed test of the LPI, violations of the quantum formulation of the LLI will result in a modulation of the contrast of the interference pattern. Violations of the classical formulation of the LLI will only modify the relative phase acquired by the amplitudes.

Finally we would like to stress that test theory including quantised internal energy of physical systems and the quantum formulation of the EEP are relevant even when internal energy operators are assumed to commute. For example, in order to describe  entanglement created between centre of mass and internal degrees of freedom, which would arise in an experiment probing WEP for a superposition of two different internal eigenstates having different accelerations of free fall due to classical violations of the WEP (see Appendix \ref{app:WEP} for more details of such a Gedanken experiment). 


 \section{Discussion}
 \label{sec:discus}
While uncontroversial in classical theory, in quantum mechanics the EEP still appears to be contentious. Nearly any claim about it can find support in the scientific literature: that the EEP is violated in quantum mechanics \cite{Davies:2004, Adunas:2001}; that there is a tension between the very formulation of the EEP and quantum theory -- and therefore EEP has to be suitably reformulated  before its validity can be discussed in quantum mechanics \cite{Laemmerzahl:1998} -- but also that  there is no difference between testing validity of the EEP in classical and quantum physics \cite{Nobili:2013}. (The latter view is motivated by the fact that so far proposed reformulations still gave rise to the  same quantitative conditions in the quantum and in the classical case. In the light of our results, this comes from the fact that such reformulations were analysed for systems with quantised only external degrees of freedom.) Below we address some of the concerns regarding the very possibility of formulating the EEP in quantum theory that are being raised.
 
Mass does not cancel from the description of a quantum system, which leads to the concern that WEP is by construction violated in quantum theory.  Note, however, that mass appears as a dimensional proportionality factor in the action of a system even in the absence of any gravitational field, both in classical and in quantum theory. While in classical theory this indeed means that the mass cancels from the dynamics of the system, it plays a physical role in quantum theory -- the mass describes how fast the phase is accumulated along the system's path and can be measured in principle even for a free particle in flat space-time (as a relative phase or via wave-packet spreading). Requiring that mass should not enter the description of a quantum system subject to gravity is therefore in contradiction with the equivalence hypothesis and thus with the rationale of the EEP. Such a requirement is tenable only in the classical limit, where only diagonal elements of the state of the system are accessible, from which the entire relative phase and thus also the mass, cancel out.

It is also often argued that for the formulation of the EEP one needs well-defined trajectories and hence point-like test systems. Since strict locality is fundamentally irreconcilable with the tenets of quantum theory (e.g.\ the uncertainty principle) it is then concluded that EEP cannot be formulated in quantum mechanics. The same argument could, however, apply also in classical statistical physics -- e.g.\ to a situation in which only the probability distribution for finding a classical particle in a certain region is known -- and hence has nothing to do with quantum physics. In practice, given a finite measurement accuracy, the relevant description of a quantum (or classical) experiment can always be restricted to a finite region. Homogeneity of gravity in such a region is then also required to hold up to a finite experimental precision. One can thus introduce a single correspondingly accelerated reference frame and meaningfully ask whether the two situations are physically equivalent. A pair of neutron experiments in which the phase shift of the interference pattern caused by gravity \cite{cow} and inertial acceleration \cite{Bonse:1983} was measured serve as just one example of successful quantum tests, in which the above conditions were met. (These two experiments together can  be regarded as a test of the non-relativistic limit of the WEP in quantum theory.)

Let us stress, that the equivalence hypothesis can be incorporated into any physical theory -- any theory can be written in arbitrary coordinates and one can postulate that the effects stemming from accelerated coordinates are equivalent to the effects of the corresponding gravitational field. In this sense the equivalence hypothesis is incorporated into quantum mechanics or quantum field theory \cite{BIR82}. Validity of the hypothesis, on the other hand, can be verified only experimentally and concerns regarding the formulation of the EEP in quantum mechanics  originate from a question: what do we need to test (and how) in order to verify whether physics in the regime where quantum effects are relevant complies with a metric picture of gravity?  Approach developed here stresses, that the very formulation of the EEP (here cited after Ref.\ \cite{clifford}) is applicable to classical and quantum theories alike, but the quantitative statement of the conditions comprising the EEP is different in the two cases. Crucial for that difference is the quantisation of the interactions, which give rise to quantised internal energies of test systems, and not the quantisation of the centre of mass alone. Most importantly, conceptual means to test conditions expressing EEP in the classical and in the quantum case in general have to be very different, since not all classical concepts apply in quantum mechanics. This, however, does not invalidate the formulation of the principle, nor by itself entails that the principle is violated.

\section{Conclusion}
\label{sec:conculsion}
We showed that due to the superposition principle of quantum mechanics the quantitative statement of the  EEP needs to be non-trivially extended in order to be applicable in quantum theory. Suitable quantum formulation of the EEP has been introduced and we showed that in order to verify its validity more parameters have to be constrained than in the corresponding classical case. Validity of the EEP in quantum theory cannot be simply inferred from classical experiments and requires conceptually new experimental approach. This provide an entirely independent merit for quantum experiments of the EEP.
Approach developed in this work can be further extended: to high energies, to incorporate possibilities of position or spin dependent mass-energies or mass-energy tensors, studied within other models. Our results shall thus largely be seen as complementary, rather than alternative, to thus far developed frameworks. 

Conceptual difference between the EEP in classical and in quantum theory pertains to the regime where quantum, special and general relativistic effects have to be jointly considered -- which has not been  accessible to experiments yet. The regime where general relativity affects internal dynamics of low-energy quantum systems seems particularly promising for the near future experimental exploration but has largely been overlooked in theoretical research. Further study of this regime can reveal new conceptual features and bring important insights into the  joint foundations of quantum mechanics and general relativity.

\begin{acknowledgments}
The authors thank I.\ Pikovski and F.\ Costa for insightful comments on the early drafts of this manuscript; and C.\ Kiefer and D.\ Giulini for  discussions. The research was funded by the Austrian Science Fund (FWF) projects W1210 SFB-FOQUS, the Foundational Questions Institute (FQXi),  and by the John Templeton Foundation.  M.\ Z.\ is a member of the FWF Doctoral Program CoQuS. 
\end{acknowledgments}

\appendix
\section{Einstein's hypothesis of equivalence; central extensions of the Galilei group and Bargmann's superselection rule.}
\label{app:conditions}
\paragraph{Einstein's hypothesis of equivalence and the EEP.} We show that the conditions derived in the main text -- imposed by the EEP on the dynamics of a massive system with internal degrees of freedom -- 
are equivalent to conditions stemming directly from requiring the validity of the Einstein's hypothesis of equivalence. We briefly discuss the relation between non-relativistic limit of the Lorentz group, central extensions of the Galilei group and mass-energy equivalence. 

As in the main text, the rest mass-energy operator of a massive system with internal degrees of freedom is denoted by $\hat M_r=m_r\hat I_{int}+\hat H_{int,r}/c^2$ and the inertial mass-energy operator by $\hat M_i=m_i\hat I_{int}+\hat H_{int,i}/c^2$. In an inertial coordinate system $(x,t)$ and in the absence of external gravitational field, the low energy limit of a Hamiltonian of such system reads
\begin{equation}
\label{schrod_eqn}
i\hbar\frac{\partial}{\partial t}=\hat M_rc^2 - \frac{\hbar^2}{2\hat M_i}\nabla ^2,
\end{equation}
 where $-i\hbar\nabla\equiv-i\hbar \frac{\partial}{\partial x}= \hat P$ is the center of mass momentum operator and where ${1}/{\hat M_i}\approx \frac{1}{m_i}(\hat I_{int}-\hat H_{int,i}/m_ic^2)$. Lorentz boost is generated by $\hat K=i\hbar t\nabla + i \hbar \frac{x}{c^2}\frac{\partial}{\partial t}$ and to lowest order in the boost  parameter $v$, the resulting new coordinates read $(x'\approx x+vt, t'\approx t+\frac{vx}{c^2})$  \cite{WeinbergQFT:1995}, thus  
\begin{equation}
\begin{cases}
\label{boost}
&\nabla = \nabla' + \frac{v}{c^2}\frac{\partial}{\partial t'}, \\
& \frac{\partial}{\partial t} = v \nabla'+ \frac{\partial}{\partial t'},\\
\end{cases}
\end{equation} 
The Hamiltonian in Eq.\ \eqref{schrod_eqn} transforms into  
\begin{equation}
\label{boost_schrod_eqn}
i\hbar\frac{\partial}{\partial t'}= \hat M_rc^2 - \frac{\hbar^2}{2  \hat M_i}{\nabla^\prime}^2 + i\hbar v \left( \frac{\hat M_r}{ \hat M_i}-1\right)\nabla' +\mathcal {O}(c^{-4}).
\end{equation}
and is invariant under the Lorentz boost if $\hat M_{i}=\hat M_{r}$.
Since the rest mass parameter $m_r$ can be assigned arbitrary value without introducing observable consequences (as long as the gravitational field produced by the system is not considered -- which is the case here), the physical requirement imposed by demanding Lorentz invariance in this limit reads $\hat H_{int,i}=\hat H_{int,r}$, as derived the main text.

Requiring  the validity of the Einstein's hypothesis of equivalence -- the total physical equivalence between laws of relativistic physics in a non-inertial, constantly accelerated, reference frame and in a stationary frame subject to homogeneous gravity  -- imposes further conditions. 
A transformation from the initial inertial frame $(x,t)$ to an accelerated coordinate system $(x''\approx x+ \frac{1}{2} gt^2, t''\approx  t+ \frac{gtx}{c^2})$, with $g$ denoting the acceleration, gives (to lowest order):
\begin{equation}
\begin{cases}
\label{transform}
&\nabla = \nabla'' + \frac{gt}{c^2}\frac{\partial}{\partial t''}, \\
& \frac{\partial}{\partial t} = gt \nabla''+\left(1+\frac{gt}{c^2} \right) \frac{\partial}{\partial t''}.\\
\end{cases}
\end{equation}
Schr\"{o}dinger equation Eq.\ \eqref{schrod_eqn} transforms under Eqs.\ \eqref{transform}  into 
\begin{equation}
\label{non-in_schrod_eqn}
i\hbar\frac{\partial}{\partial t''}= \hat M_rc^2- \hat M_rg x + i\hbar gt \left( \frac{ \hat M_r}{ \hat M_i}-1\right)\nabla'' - \frac{\hbar^2}{2 \hat M_i}{\nabla''}^2.
\end{equation}
For a massive particle subject to a homogeneous gravitational potential $\phi(x)=gx$ its coupling to gravity is given by its gravitational charge -- the total gravitational  mass-energy $\hat M_g=m_g\hat I_{int}+\hat H_{int,g}/c^2$, where $m_g$ describes the gravitational mass parameter and $\hat H_{int,g}$ contribution to the mass from internal energy. The Hamiltonian of such a system reads
\begin{equation}
\label{grav_schrod_eqn}
i\hbar\frac{\partial}{\partial t'}=\hat M_rc^2-\hat M_g g x - \frac{\hbar^2}{2 \hat M_i}{\nabla'}^2.
\end{equation}
Thus, for the validity of the Einstein's Hypothesis of Equivalence 
in addition to $\hat H_{int,i}=\hat H_{int,r}$ it is also required that $\hat M_g=\hat M_i$ -- in full agreement with the derivation in the main text. Moreover when the hypothesis of equivalence  holds, the Hamiltonians of a composed quantum system subject to weak gravity reduces to the Hamiltonian in Eq.\ \eqref{meep_ham}. 

\paragraph{Central extensions of the Galilei group and  Bargmann's superselection rule.} When Lorentz invariance holds $\hat M_r=\hat M_i\equiv\hat M$ and in the low energy limit $i \hbar \frac{1}{c^2}\frac{\partial}{\partial t}\approx  \hat M$. The boost generator takes the form $\hat K\approx i\hbar t\frac{\partial}{\partial x} + x \hat M$ This is a boost generator of a central extension of the Galileo group with central charge $\hat M$: (see e.g.\ Refs.:\cite{WeinbergQFT:1995, Bargmann:1954, Levy-Leblond:1963, Giulini:1996}) - i.e.\ a group which quotient by the one-parameter subgroup generated by $\hat M$ is isomorphic to the Galilei group, and where the additional generator commutes with all others. This can be seen from the commutator of $\hat K$ and $ \hat P$ -- the generator of translations -- which here reads: $[\hat K, \hat P ]=i\hbar \hat M$, whereas it vanishes for the Lie algebra elements of the Galilei group. Non-vanishing of the above commutator is a direct consequence of the fact that the non-relativistic limit of the Lorentz group describing a particle with a mass parameter $m$ is the central extension of the Galilei group with central charge $m$, not the Galilei group \cite{WeinbergQFT:1995, Inoenue:1952}. 

On the other hand, boost generator of the physical representation of the Galilei group on the state-space of a non-relativistic particle with mass $m$ reads $\hat K'= i\hbar t\frac{\partial}{\partial x} + mx$.
Therefore, generators of the physical representation of the Galilei boost and shift also do not commute: $[\hat K', \hat P ]=i\hbar m\hat I$.
This results in a mass-dependent phase factor in transformations of physical states under Galilei group, which means that the non-relativistic quantum theory admits a projective representation of the Galilei group, rather than a proper representation.  As long as the mass is just a parameter of the theory, this is just  an unobservable global phase.  Unitary representations of the central extensions of the Galilei group with the central charge being a parameter  and projective representations of the Galilei group are physically fully equivalent. 
Considering that mass, like other physical observables, could in principle be an operator with different eigenvalues, lead to the formulation of the superselection rule \cite{Bargmann:1954} under the assumption that the ``real'' symmetry group of the non-relativistic theory is the Galilei group. We recall the argument below. Let us denote by $g$ and $g'$ the Galilei group elements of a spatial translation by $a$ and a boost by $v$, respectively. They satisfy: $g^{\prime -1} g^{-1} g' g = 1$ (identity element the Galilei group). However, the Hilbert space representation of these transformations, implemented by unitary operators $\hat U_g$, $\hat U_{g'}$, satisfies $\hat U_{g'}^{-1}\hat U_{g}^{-1}\hat U_{g'}\hat U_{g}=e^{-imva}\hat I$.
Applying this combination of boosts and translations to a superposition state of masses $m$ and $m'$ would result in a relative phase $e^{iva(m-m')}$ and therefore a different physical state, unless $m = m'$. However, this operation shall represent the identity transformation of the Galilei group and cannot alter physical states. Hence a superposition of states with different masses is considered unphysical in a theory with Galilei invariance and is ``forbidden'' -- this is the original argument of Bargmann behind  the superselection rule for the mass. 

It has been noted in e.g.\ Refs.\ \cite{Giulini:1996, Greenberger:2001}, that  considering superpositions of states with different masses is only consistent in a theory where mass is a an operator  $\hat m$ -- a generator of shifts of its conjugate, new degree of freedom. Non-relativistic quantum as well as classical theory with a dynamical mass admits central extension of the Galilei group as a symmetry, and not the Galilei group -- there is no ambiguity in this case any more.
Moreover, for the central extension of the Galilei group the above analysed combination of the shift and boost elements, which for the central extensions we denote by $\tilde g$ and $\tilde g'$,  satisfies $\tilde g^{\prime -1} \tilde g^{-1} \tilde g' \tilde g = \tilde g''$ where $\tilde g''$ is an element of the central extension of the Galilei group generated by the mass operator, which shifts the  degree of freedom conjugate to the mass by $va$. The unitary representation of this operations on the Hilbert space via operators $\hat U_{\tilde g}$, $\hat U_{\tilde g'}$, satisfies $\hat U_{\tilde g'}^{-1}\hat U_{\tilde g}^{-1}\hat U_{\tilde g'}\hat U_{\tilde g}=e^{-i\hat mva}=\hat U_{\tilde g''}$. Thus,  the non-relativistic theory with mass treated as dynamical degree of freedom admits a proper representation of the central extension of the Galilei group.  This, however, means that there is no need for a superselection rule for the mass. The question thus arose whether one can or cannot superpose states with different masses in the non-relativistic theory? If that is not possible -- what is the dynamical reason for that and what is the dynamical meaning of the superselecton rule? (See outlook  in  Ref.\ \cite{Giulini:1996}.)  Approach presented in this work shows that such non-trivial central charge has a natural physical interpretation as a mass-energy operator of a system with internal degrees of freedom.  Non-trivial central extensions of the Galilei group, both in quantum and in classical physics, can be seen as an describing  relativistic, point-like systems with internal dynamics in the low-energy, but not fully non-relativistic, limit. Internal energy effectively contributes to the mass, rendering it to be dynamical, and drives dynamics of the internal degrees of freedom. 
 Such an approach is fully consistent with the low-energy limit of the Poincar\'e group and with the observation made in \cite{Greenberger:1974} (see also references therein) that treating mass as a dynamical variable in a non-relativistic theory introduces the relativistic notion of proper time and time dilation effects. Approach proposed here, however, does not require introducing  any new  degrees of freedom -- the effective mass-energy operator acts on internal states, such as vibrational or electromagnetic energy levels of atoms, molecules, etc. 
 
The above observations allow to understand the result of Bargman, that superpositions of states with different masses are non-physical in the non-relativistic limit,  without postulating any superselection rule and within a mathematically consistent approach. Taking the non-relativistic limit only for the external degrees of freedom of a composite, relativistic system yields a system described by a central extension of the Galilei group, with the central charge given by the dynamical mass-energy operator $\hat M$ describing evolution of the internal degrees of freedom. Such a theory still features time dilation effects on the internal evolution, but a consistent non-relativistic limit has to give rise to the  non-relativistic, Euclidean, space-time, with absolute time. This is the case if the central charge in the low-energy regime has a general structure $\hat M\approx m\hat I+ \mathcal{O}(1/c^2)$. From this perspective, the result that in the fully  non-relativistic limit no superpositions of different masses are allowed is simply a consequence of a consistent, operational definition of a non-relativistic theory. 


\section{Fully classical test theory of the EEP}
\label{app:classical}
In classical physics Hamiltonian of a composite system is a function of phase space variables of the centre of mass $(Q,P)$ and of the internal degree of freedom $(q, p)$  with the internal mass-energies  $M_\alpha=m_\alpha c^2+E_\alpha$ and reads
\begin{equation}
\tilde H^C_{test}= M_r+\frac{P^2}{2M_i}+M_g\phi (Q) \approx m_rc^2+E_r+\frac{ P^2}{2m_i}+m_g\phi ( Q)-E_i\frac{P^2}{2m_ic^2}+E_g\frac{\phi(Q)}{c^2}.
\end{equation} 
Time evolution of a classical variable is obtained from its Poisson bracket with the total Hamiltonian: $d/dt=\{\cdot,\tilde H^C_{test}\}_{PB}$.
The acceleration of  the center of mass $Q$ reads 
\begin{equation}
\label{uff}
\ddot Q = -{M_g}{M_i}^{-1} \nabla \!\phi(Q)\;,
\end{equation}   
where $\nabla$ is derivative with respect to $Q$. Eq.\ \eqref{uff} recovers the result that free fall is universal if $M_g={M_i}=1$ (or more generally, $M_g /{M_i}$ can be any positive number, the same for all physical systems, but such a numerical factor would just redefine the gravitational potential). 

The time evolution of the internal variable $q$ (keeping only first order terms in $H_{int, \alpha}/m_\alpha c^2$) reads
\begin{equation}
\label{time_dilation}
\dot q(Q,P) = \dot q_r-\dot q_i \frac{P^2}{2m_i^2c^2}+\dot q_g\frac{\phi(Q)}{c^2},
\end{equation}
where $\dot q_\alpha:=\{q, H_{\alpha}\}_{PB}$ are in principle different velocities. 
The gravitational time dilation factor $\Delta \dot q/\dot q := \frac{\dot q(Q+h,P) - \dot q(Q,P)}{\dot q(Q,P)}$ reads
\begin{equation}
\label{cgr}
\Delta \dot q/\dot q \approx \frac{\dot q_g}{\dot q_r}\frac{\nabla \!\phi(Q) h}{c^2},
\end{equation}
and it reduces to that  predicted by general relativity $\Delta \dot q/\dot q \approx \frac{\nabla \!\phi(Q) h}{c^2}$  if $H_{int,r}=H_{int,g}$. Similarly, universality of special relativistic time dilation is recovered  if $H_{int,r}=H_{int,i}$. 

Conditions for the validity of the EEP (and the number of parameters to test) are the same in the fully classical case above and in the model $H^C_{test}$ which describes a system with quantised centre of mass degrees of freedom. Since the EEP imposes equivalence conditions on the mass-energies of the system, it is the quantisation of the internal energy which is relevant for the difference between the classical and the quantum formulation of the EEP. 


\section{Lagrangian formulation of the  test theory}
\label{app:lagrangian}
Lagrangian formulation of the test theory is obtained from the Legendre transform of the test Hamiltonian. The derivation is valid for both the classical and the quantum model; we will thus write for brevity $H_{test}= m_rc^2+H_{int,r}+\frac{ P^2}{2m_i}+m_g\phi(Q) - H_{int,i}\frac{P^2}{2m_i^2c^2}+H_{int,g}\frac{\phi(Q)}{c^2}$. 

For the centre of mass degree of freedom the canonically conjugate velocity is given by 
\begin{equation*}
\dot {Q}=\frac{\partial H_{test}}{\partial  P} = \frac{ P}{ m_i}\left(1-\frac{ H_{int,i}}{m_ic^2}\right).
\end{equation*}
We formally introduce position $q$ and momentum $p$ of the internal degrees of freedom, which dynamics is given by the  Hamiltonians $H_{int, \alpha}=H_{int, \alpha}(q,p)$. The conjugate internal velocity is thus defined as $\dot q= \frac{\partial H_{test}}{\partial p}$ and reads
\begin{equation*}
\dot {q}=\frac{ \partial H_{int,r}}{ \partial p}- \frac{ \partial H_{int,i}}{ \partial p} \frac{P^2}{2m_i^2c^2}+\frac{\partial H_{int,g}}{\partial p}\frac{\phi(Q)}{c^2}.
\end{equation*}
Lagrangian of the test theory can now be obtained through the Legendre transform of $H_{test}$:  $L_{test}:=P\dot Q+p\dot q-H_{test}$. We first introduce the total internal Lagrangians $L_{\alpha}$ via the Legendre transform of the total internal mass-energies $m_\alpha c^2+H_{int,\alpha}$: 
\begin{equation*}
L_{\alpha}:=\frac{\partial H_{int}}{\partial p}p-m_\alpha c^2-H_{int, \alpha}\equiv -m_\alpha c^2+L_{int,\alpha}, 
\end{equation*}
which  leads the test Lagrangian in the form:
\begin{equation}
\label{ltest}
L_{test}= L_{r}-L_{i}\frac{\dot Q^2}{2c^2}+L_{g}\frac{\phi(Q)}{c^2}.
\end{equation}
Note, that $-m_\alpha c^2$ is the non-dynamical part of the internal Lagrangian and  $L_{int,\alpha}$ is its dynamical part -- in a full analogy to the Hamiltonian picture where $mc^2$ is the non-dynamical and $ H_{int,\alpha}$ the dynamical part of the internal mass-energy. The conditions for the validity of the EEP derived in the main text for the internal Hamiltonians now translate to $L_i=L_r=L_g$. Indeed, when the internal dynamics is universal $L_\alpha\equiv L_0$ the Eq.\ \eqref{ltest} reduces to
\begin{equation}
\label{ltest_lim}
L_{test}\xrightarrow{L_\alpha\equiv L_0}L=L_{0}(1-\frac{\dot Q^2}{2c^2}+\frac{\phi(Q)}{c^2}).
\end{equation}
Eq.\ \eqref{ltest_lim} is the lowest order approximation to the dynamics of a particle in space-time given by e.g.\ the Schwarzschild metric. Indeed,  $L\approx L_0\sqrt{-g_{\mu\nu}\dot x^\mu\dot x^\nu}$ with metric elements $g_{00}\approx-(1+2\phi(x)/c^2)$, $g_{ij}\approx c^{-2}\delta_{ij}$, $g_{0i}=g_{i0}=0$ $i,j=1,2,3$. In the limit $L_0\approx -mc^2$ the  non-relativistic Lagrangian of a massive particle in Newtonian potential is recovered $L\approx-mc^2+m\dot Q^2/2-m\phi(Q)$.  

In contrast to thus far considered  test theories of the EEP for composed systems, which only incorporate internal (binding) energy as  fixed parameters, test theory given by the Lagrangian in Eq.\ \eqref{ltest}  incorporates the dynamics of the associated degrees of freedom. 

\section{Quantum test of the classical WEP}
\label{app:WEP}
Assume that WEP holds but only in the Newtonian limit, $m_i=m_g\equiv m$,  and that LLI is valid ($\hat H_{int,r}=\hat H_{int,i}$) but $\hat H_{int,i}\neq\hat H_{int,g}$. In particular, we restrict to classical violations of the WEP, i.e.\ $[\hat H_{int,i},\hat H_{int,g}]=0$. For an internal energy eigenstate  $|E_j\ket$ we have $\hat M_{i}|E_j\ket=M_{1,i}|E_j\ket$ and $\hat M_{g}|E_j\ket=M_{j,g}|E_j\ket$ where $M_{j,\alpha}=m+E_{j,\alpha}/c^2$. From Eq.\ \eqref{q_uff} we obtain $\ddot {\hat Q }|E_j\ket= -g_j|E_j\ket$ ($j=1,2$) where $g_j=gM_{j,g}/M_{j,i}$ where we assumed homogeneous  gravitational field $g$. Parameters describing possible violations are $\eta_j:=M_{j,g}/M_{j,i}$ (which can be seen as the diagonal elements of the matrix $\hat \eta$ introduced in the main text). When $\eta_1\neq \eta_2$ for some two internal states $|E_1\ket$, $|E_2\ket$ the centre of mass will have the free-fall acceleration that depends on the internal state. Consider now a coherent superposition of the two internal energy eigenstates,  semi-classically localised at some height $h$: 
\begin{equation}
\label{psi0}
|\Psi(0)\ket=1/\sqrt{2}(|E_1\ket+|E_2\ket)|h\ket. 
\end{equation}
Under free-fall it evolves into 
\begin{equation}
\label{psit}
|\Psi(t)\ket=1/\sqrt{2}(e^{i\phi_1}|E_1\ket |h_1\ket+e^{i\phi_2}|E_2\ket|h_2\ket),
\end{equation}
where $h_j=h-1/2 g_j t^2$, $j=1,2$ is the position of the centre of mass correlated with the internal state $|E_j\ket$ after time $t$ of free fall and $\phi_{j}(t)$ is the free propagation phase for a particle with a total mass $M_{j,i}$ under gravitational acceleration $g_j$, which can be found e.g.\ in \cite{Storey:1994}. Initial superposition in a presence of classical violations evolves into an entangled state, with the internal degree of freedom entangled to the position. As a result the reduced state of the internal degrees of freedom $\hat \rho_{int}(t)$ becomes mixed: $\hat \rho_{int}(t):=\textrm{Tr}\{|\Psi(t)\ket\bra\Psi(t)|\}=1/2(|E_1\ket\bra E_1|+|E_2\ket\bra E_2|+e^{i\phi_1-i\phi_2}\bra h_2|h_1 \ket|E_1\ket\bra E_2| +h.c)$. The amplitude of the off-diagonal elements 
\begin{equation}
\label{visib_class}
\mathcal{V}:=|\bra h_2|h_1 \ket|
\end{equation}
quantifies the coherence of the reduced state and it decreases with the position amplitudes becoming distinguishable, in agreement with the quantum complementarity principle for pure states, see e.g.\ \cite{englert}. When the position amplitudes become orthogonal we have $\mathcal{V}=0$  and the reduced state becomes maximally mixed. The classical violations of the WEP and the superposition principle of quantum mechanics  thus entail  decoherence of any freely falling system into its internal energy eigenbasis. 

Since we assumed the validity of the LLI but a violation of the WEP we shall also observe a related violation of the LPI.  Indeed, a coherent superposition of different energy states evolves in time and thus constitutes a ``clock''. A frequency of such a ``clock'' is given by the inverse of the energy difference between
the superposed states. The internal state in Eq.\ \eqref{psi0} when trapped at a height $h$ evolves in time at a rate 
 $\omega(h)=\omega(0)(1+(E_{2,g}-E_{1,g})/(E_{2,i}-E_{1,i})gh/c^2)$ where $\omega(0)=(E_{2,i}-E_{1,i})/\pi\hbar$, in violation of the LPI. In case of no violations this rate would read $\omega(h)_{GR}=\omega(0)(1+gh/c^2)$.  An anomalous frequency dependence on the system's position in the laboratory frame $\omega(h)$  would be the only consequence of the classical violations of the LPI for classical clocks.  However, for a quantum ``clock'' there is an additional effect: The final state of the internal degree of freedom in Eq.\ \eqref{psit} is  stationary (because it becomes fully mixed). Classical violations discussed above thus result in a decoherence of any time evolving state, a ``clock'' into a stationary mixture.

Decoherence effect and entanglement between internal and external degrees of freedom, that would arise as a result of the classical violations of the WEP, cannot be described within a fully classical theory. Quantum test theory of the EEP is therefore necessary in order to describe all effects of the EEP violations on quantum systems, even if the violations themselves are assumed to be classical.

Realisation of such a quantum test of the classical WEP in principle takes place in interferometric experiments where atoms propagating in the two arms of the interferometer are in different energy eigenstates (Raman beam-splitting). As an example we consider a recent experiment performed by the group of P.\ Bouyer \cite{Geiger:2011}. In this experiment  Mach-Zehnder interferometer with $^{87}$Rb  was operated during a ballistic flight of an airplane with the aim to provide a proof of principle realisation of an inertial sensor in microgravity. 
We approximate the centre of mass position of the atoms by a Gaussian distribution $\bra x|h_j\ket \propto  e^{-(h_j-x)^2/2l_c^2} $ where $l_c$ is the coherence length of the atom's wave-function. Assuming small violations the visibility in Eq.\ \eqref{visib_class} can be approximated to $\mathcal{V} \approx 1-(\Delta \eta \frac{gT^2}{l_c})^2$, where $\Delta \eta=|\eta_1-\eta_2|$. From the experimental parameters estimated in \cite{Geiger:2011}: $\mathcal{V}\approx 0.65$, $T=20$ ms and estimating $l_c\approx 10$ $\mu$m we can infer a bound $\Delta \eta<8 \cdot 10^{-3} $.

\bibliographystyle{linksen}
\bibliography{bibliomz}

\end{document}